\documentclass[aps,twocolumn]{revtex4}
\usepackage{graphicx}
\usepackage{dcolumn}
\begin{document}
\def\be{\begin{equation}}
\def\ee{\end{equation}}
\def\lag{\langle}
\def\rag{\rangle}
\title{Equilibrium correlations in a model for multidimensional epistasis} 
\author{G\"une\c s S\"oyler$^1$ and
Ay\c se Erzan$^{1,2}$}
\affiliation{$^1$  Department of Physics, Faculty of  Sciences
and
Letters\\
Istanbul Technical University, Maslak 80626, Istanbul, Turkey }
\affiliation{$^2$  G\"ursey Institute, P. O. Box 6, \c
Cengelk\"oy 81220, Istanbul, Turkey}

\date{\today}
\begin{abstract}

We investigate a statistical model for multidimensional epistasis. The
genotype is devided into subsequences, and within each subsequence
mutations which occur in a prescribed order are beneficial.  The
bit-string model used to represent the genotype, may be cast in the form
of a ferromagnetic Ising model with a staggered field. We obtain the
actual correlations between mutations at different sites, within an
equilibrium population at a given {\it tolerance}, which we define to be
the temperature of the statistical ensemble.


\end{abstract}
\maketitle
\section{Introduction}
Although evolution takes place via a combination of random mutations and natural selection, it seems to proceed rather rapidly along directed paths in the space of all possible genetic states. It is a challenging problem to try to understand the mechanisms which lead to this phenomenon~\cite{Smith}. 

Eigen has pointed out that each ``species'' actually consists of a more or less narrow distribution in the phase space of all possible genetic states, and this distribution may shift, in response to environmental pressure~\cite{Eigen}.  Natural selection in response to environmental factors is usually modelled in terms of a ``fitness function'' which is a measure of the survival probability and/or reproductive capability of the individual.  

Those mutations  which have a salutary effect on the fitness persist in the population and lead to new variants; other, neutral  mutations may simply be carried along since they do not affect the well being of the individual. Deleterious mutations usually affect the organism adversely, and the accumulation of too many will reduce the fitness drastically.

The simplest hypothesis biologists have adopted regarding how the number
of mutations affect the fitness, is that each deleterious mutation reduces
the fitness by an identical factor, say $1/a$, $a > 1$.  This is
equivalent to assuming that the effect of each deleterious mutation is
independent of the others, or that there is no ``epistasis'' between the
mutations, and leads to a fitness function which decays exponentially with
$m$, the number of mutations, as $f \sim \exp(-\alpha m)$, where $\alpha =
\ln a$.~\cite{Kondrashov1} A different type of assumption can be made, to
take $f$ to depend on $m$ in a step-wise fashion, so that the value of $f$
is unaffected for $m$ less than a threshold, after which it is reduced
drastically.~\cite{Jan}

It is clear, however, that there can be epistatic interactions between
mutations at different points on the genetic string and that the
expression of unmutated genes may be affected by the presence of mutations
at certain loci, and so on.~\cite{Book} Therefore $f$ may depend not only
on the total number of mutations, but also on their location, and may also
increase as the result of mutations at certain loci.  It has recently been
pointed out that the fitness may depend strongly on the order in which
certain mutations may occur~\cite{Kondrashov}.  As a case in point, for a 
mutation
leading to a certain modification to be beneficial, one must already have
had a mutation leading to the emergence of a feature which will benefit
from this modification.

This type of epistasis actually lends itself to a treatment in terms of
statistical equilibria, with the appropriate choice of a fitness 
function.~\cite{Seher}

In this paper we will represent a complete genomic sequence with epistatic
interactions by a one dimensional feromagnetic Ising model. We will
subdivide the total genotype into subsequences (here taken to be of length
2, without any loss of generality), and stipulate that mutations can lead
to salutory effects only if they occur in a certain order within these
subsequences. We will further introduce a new quantity, the ``tolerance''
of the environment, which will have to be taken into acount to determine
how strongly epistatis interactions affect the overall fitness. Our aim
will be to compute, within this model, the effective correlations between
mutations at different sites, at fixed tolerance, within a population at
equilibrium.

\section{The Model}

Since Eigen first introduced the quasi-species model~\cite{Eigen}
bitstring models of genetic evolution have been extensively studied
numerically~\cite{Oliveira,Orcal,Tuzel1,Tuzel2}. In this approach, the
genotype of an individual is represented by a string of Boolean variables
$\sigma_i$, $i=1,\ldots N$, which can obviously be identified with a one
dimensional system of Ising spins~\cite{McCoy}. If one takes the wild
type, or the initial genotype, to consist of a string of 0's, each point
mutation is indicated by flipping the bit representing a given gene, from
0 to 1.

We would like to avail ourselves of the analytically known results on the
exactly solvable Ising model in equilibrium, to be able to make definite
predictions regarding the correlation of mutated genes on a given
genotype, under assumptions similar to those of Kondrashov and
Kondrashov~\cite{Kondrashov}.

We devide the one dimensional string of spins representing the state of
the genome, into dimers. We demand that the fitness is only increased
relative to the wild type (all zeroes) if the bits that flip to 1 occur
sequentially. \cite{Kondrashov}. Thus, within each dimer, $(0,0)$,
$(1,0)$, $(1,1)$ are in increasing order of fitness while $(0,1)$ is less
fit than $(0,0)$.

Let us first construct a cost function by defining the Ising Hamiltonian,
\be 
{\cal H}= -J/2 \sum_i s_i s_{i+1} - K\sum_{i {\rm odd}} s_i - H \sum_{i {\rm even}}s_i \;\;\;, \label{Ising}
\ee
where for greater convenince in manipulation, we have defined the variables $s_i= 2(\sigma_i - 1/2)$. The value of ${\cal H}$ for each given sequence of $\{s_i\}$ will serve as a cost function, in terms of which we may define the fitness.  
Notice that in the first term, we have a coupling between nearest neighbors, which tends to reduce the ``cost'' for those configurations in which the adjacent ``spins''are in the same state.  If the  constants,  $K$ and $H$, which correspond to a staggered  external field in an Ising model, are here chosen as $K=3J/4$ and $H=-J/4$, then we obtain a situation in which the dimer configurations  $(-1,1)$, $(-1,-1)$, $(1,-1$ and $(1,1)$ have decreasing cost.

Then $f$ is defined as 
\begin{equation}
f = {1 \over Z}  e^{-\beta H} \label{Boltz}
\end{equation}

where $\beta$ is a measure of how effective the cost function is in
affecting the fitness, and $Z$ is a normalization factor so that $f \in
(0,1)$. Note that $f[\{s_i\}]$ can be identified as the Boltzmann factor
in an equilibrium statistical model with the Hamiltonian $H$, at constant
inverse ``temperature" $\beta^{-1}$, and corresponds to the probablity of
observing, within an equilibrium population, the particular genotype
$\{s_i\}$. Temperature may be seen as the amount of randomness, or
disorder in the system, competing with the cost function in determining
the fitness.  The higher the temperature, or randomness, the weaker will
be the effect of the cost function in determining the state of the system.
Therefore we define
\be T\equiv\beta^{-1}\ee
as the {\it tolerance} in the system. Here $J$ is a measure of the
strength of the interaction between the states of each of the sites
(alleles), $\sigma_i$.  Clearly, $\beta$ and $J$ will always occur
together in this model, in the product $\beta J$, and we may simply absorb
$J$ into the definition of $\beta$.
         
The fitness $f$  is normalized to take values between $(0,1)$, by defining 
\begin{equation}
Z\equiv \sum_{\{s_i\}} e^{-\beta H[\{s_i\}]}\;\;\;.
\end{equation}
Using the transfer matrix method, this sum may be computed exactly. 
We may then compute the expectation values $m_i=\langle s_i\rangle$.  
Note that the quantity $(m_i+1)/2$ corresponds to the probability of
finding a mutation on either of the sublattices, $i$ odd, or $i$ even.  
The results are shown in Fig. 1, as a function of $T/J$, which is the
(inverse) ration of the strength of the epistasis to the tolerance in the
system.  The ``staggered magnetization,'' $m_s \equiv \langle s_{i\, {\rm
(odd)}} - s_{i\,{\rm (even)}} \rangle$ is shown in Fig. 2, and is twice 
the 
difference
between the probabilities of encountering a mutation on either of the two
sublattices (the first or the second sites beloging to a dimer).  It is
seen to peak sharply at small values of the tolerance, and then the
difference decays to zero, as the tolerance becomes very large, at which
point the fitness function becomes essentially flat.

In Fig. 3a, b and c, we display the correlation functions, $C_2 = \langle
s_i s_{i+2}\rangle$ and $C_1=\langle s_i s_{i+1}\rangle$, as well as the
subtracted correlation function $C_s \langle s_i s_{i+2}\rangle - \langle
s_i\rangle \langle s_{i+2}\rangle$, as a function of $T/J$.  It can
clearly be seen here as well, that the effect of epistatic interactions in
building up correlations between mutated sites on the gene string, is felt
strongly within a given range of tolerances, in units of the strength of
interaction.  At $T=0$, since the genes are in the ordered state with all
$s_i=1$, the excess correlation $C_s$ due to the interactions, is nil.  
In the other extreme of very large tolerances, the system is completely
disordered, correlations vanish, so that the two terms in $C_s$ tend to
each other, and both tend to zero.

To further elucidate the meaning of tolerance, we may compute the relative
variances $v_m$, where

\be
v^2_m \equiv \langle (s_i  - m_i)^2\rangle \;\;\;.
\ee
It is easy to see, that within a mean field approximation, where all the spins interact pairwise with each other, i.e., ${\cal H} = -J/N \sum_{(ij)}s_is_j$, $v_m^2 = T/2J$; thus the ratio $T/J$ is a measure of the size of the fluctuations about the mean.  In genome space, this means $\sqrt{T/J}$ is a measure of the radius of the distribution of genotypes about the most frequently encountred one, in equilibrium.

\section{conclusions}

In summary, we have cast an epistatic quasispecies model interms of a one
dimensional Ising model with staggered magnetic field, to give greater
advantage to certain subsequences of genes that may be mutated.  We
defined a ``tolerance'' of the system, to introduce an equilibrium
statistical ensemble, namely one whose statistical properties do not
change in time.  Correlations induced on the genetic sequence of
individuals in this equlibrium population have been computed as a function
of the tolerance and the strength of the epistatic interaction, using
exact solutions of the Ising model in one dimension.  It has been shown
that non-trivial correlations between mutated sites on the gene string may
arise only in a finite range of the tolerance for a given interaction
strength.

{\bf Acknowledgements}

We are grateful to Seher \"Oz\c celik for many intersting discussions.  AE acknowledges partial support from the Turkish Academy of Sciences.



{\bf Figure captions}

1. The magnetization 
 at a) odd, b) even 
sites,  of the one dimensional 
Ising model on these respective sublattices, as a function of the 
``tolerance." The probability of encountering mutations at these 
respective sites is given by $(m_i+1)/2$.

2. The ``staggered magnetization" is twice the difference between the 
probabilities of encountereing mutated genes at the first or the second 
site of the dimers into which the genome has been decomposed.

3. The correlations function between mutated sites on a) analogous sites
on neighboring dimers, b) odd-even sites c) the subtracted correlation 
function between analogous sites.  

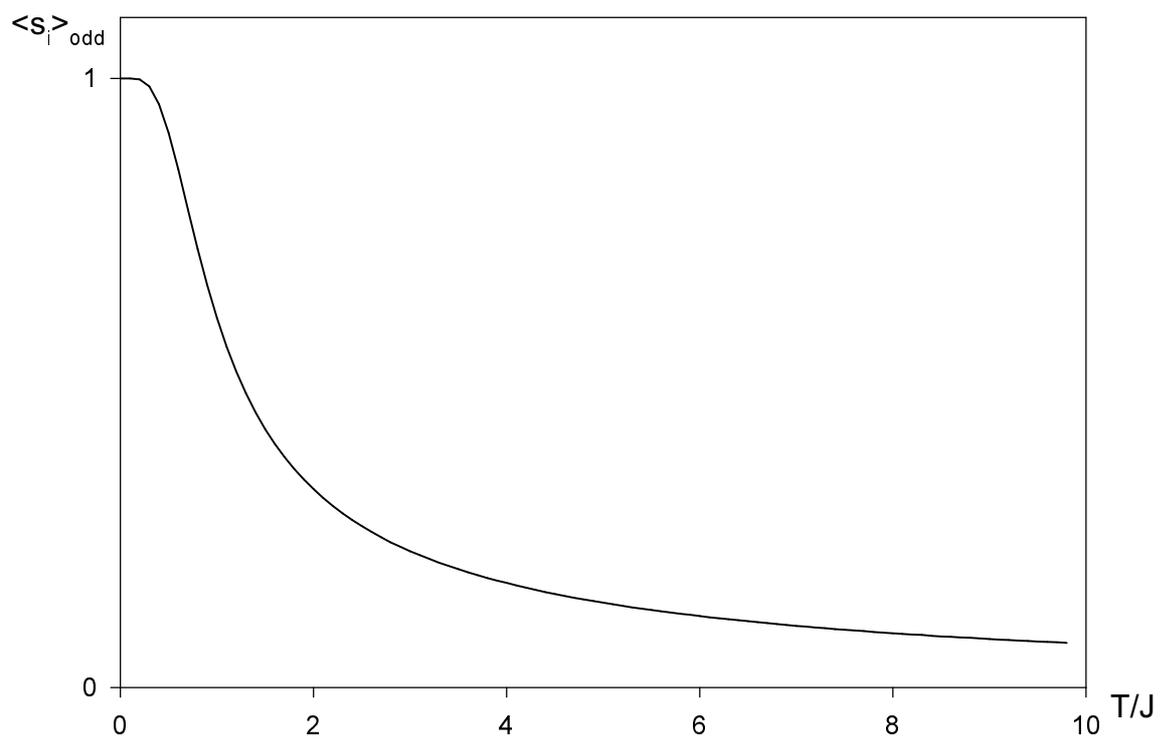
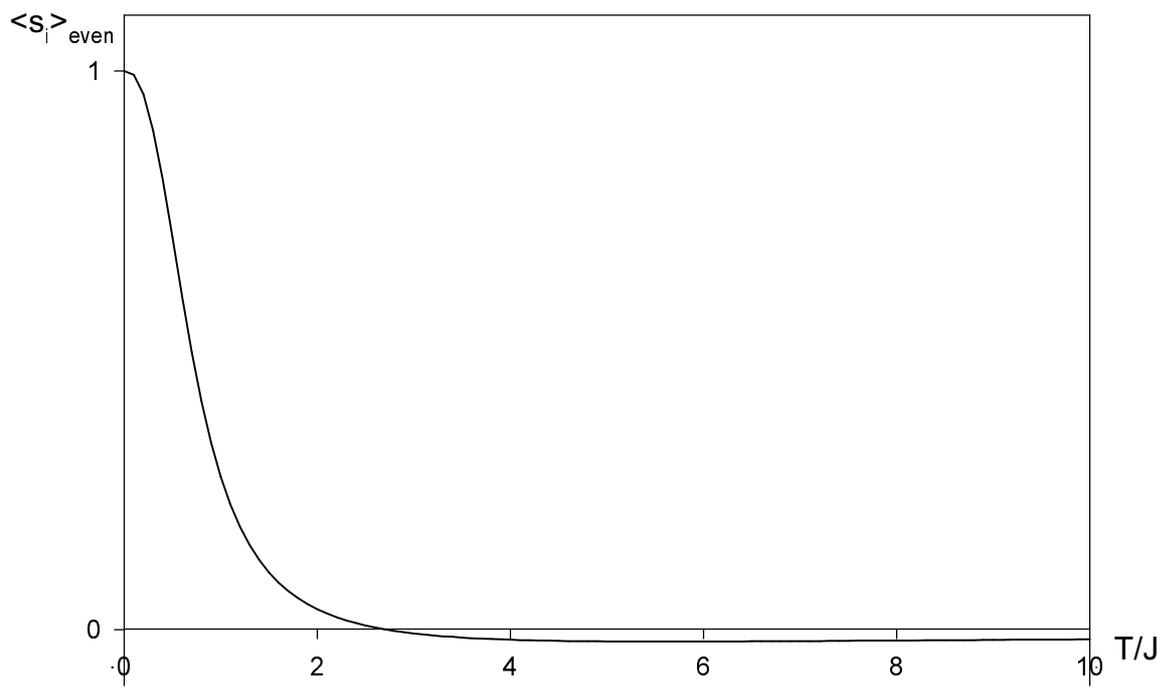
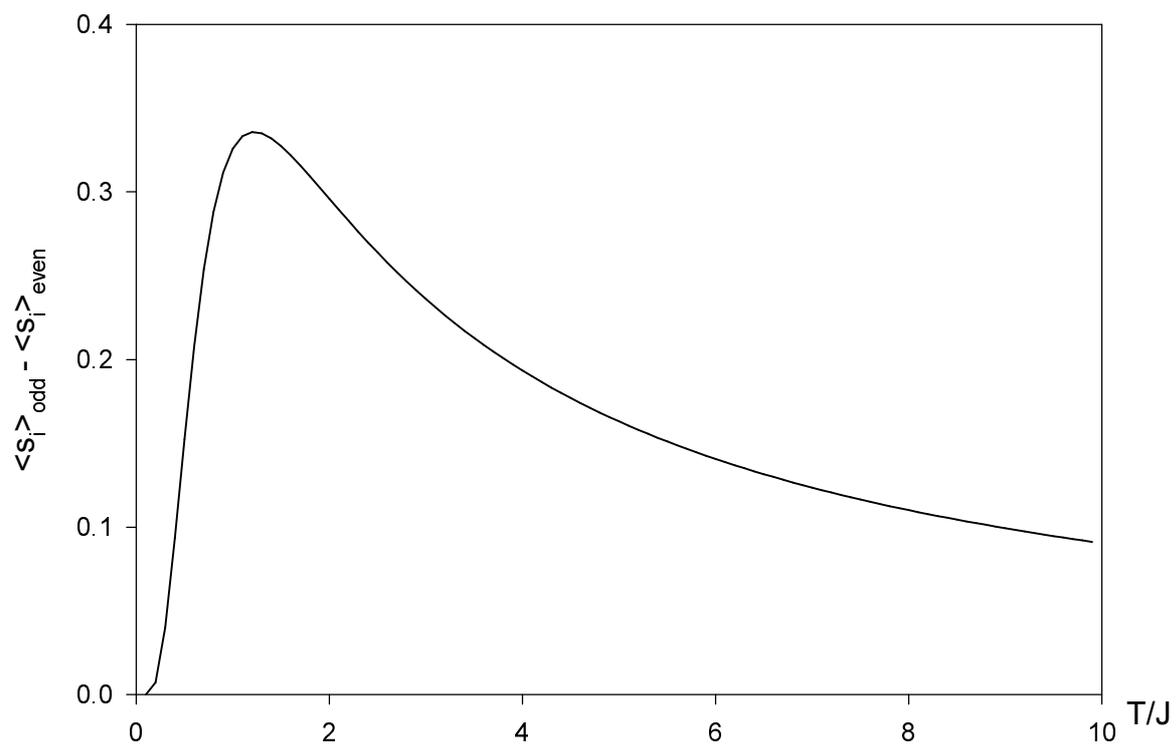
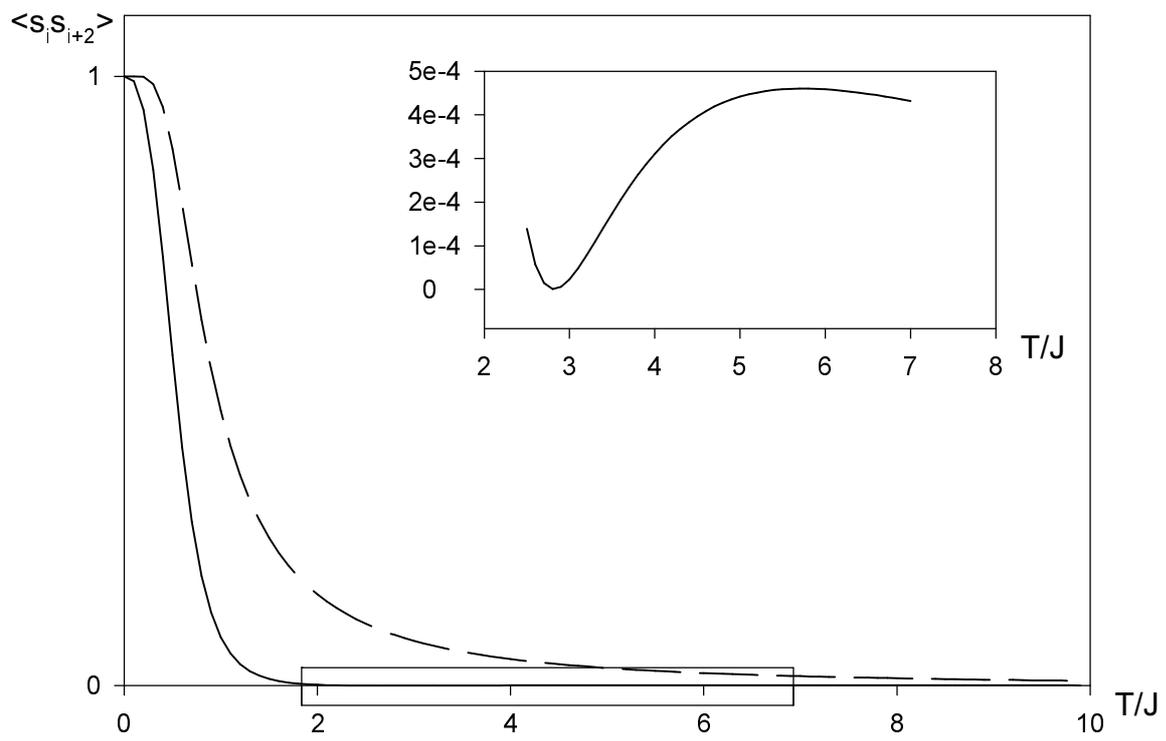
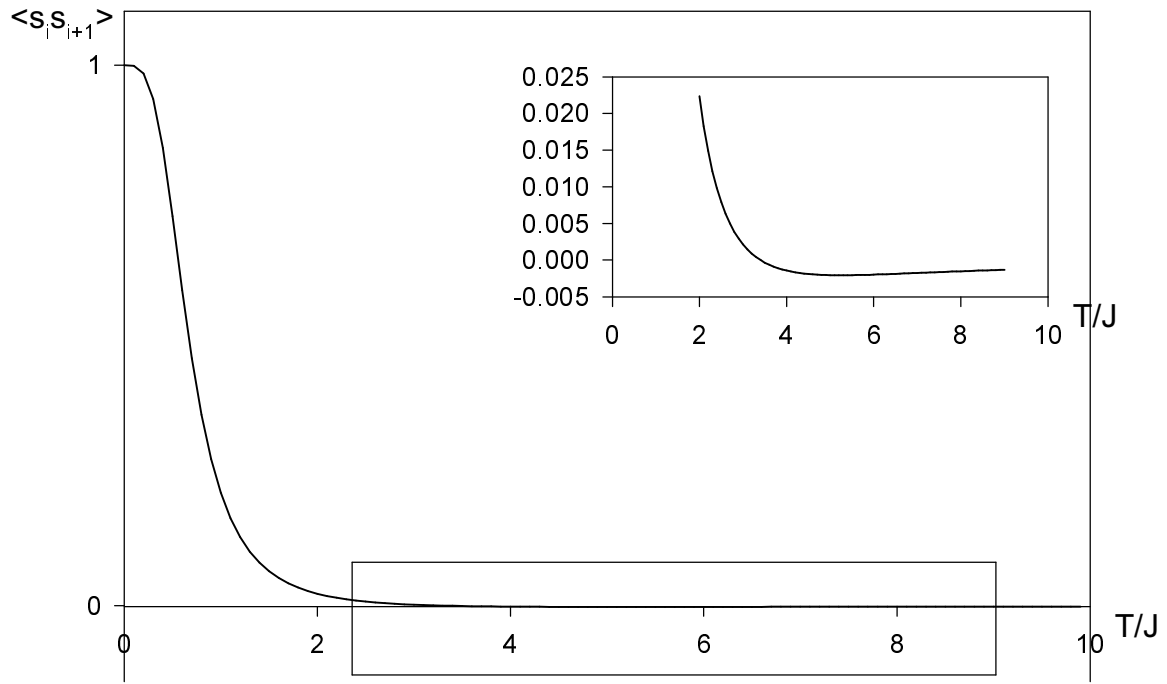
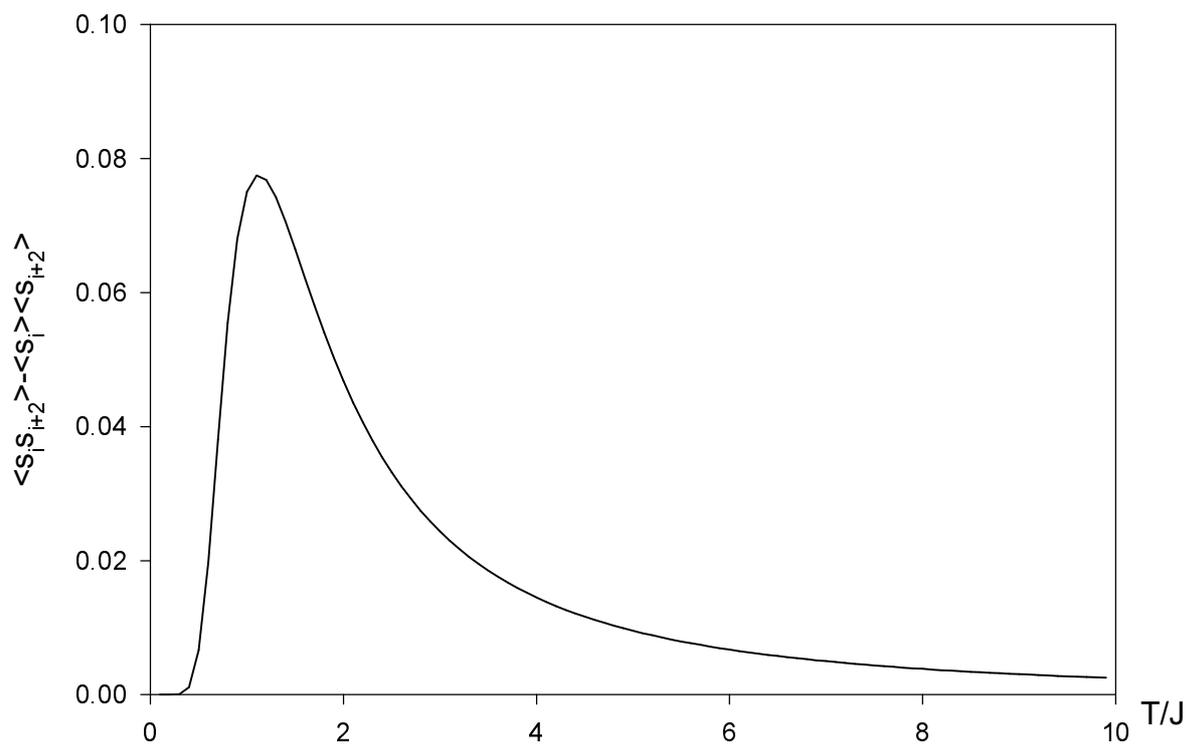
\end{document}